\documentclass[conference]{IEEEtran}
\usepackage{amssymb}
\usepackage{color}
\usepackage{multirow}
\usepackage{mathrsfs}
\newtheorem{theorem}{Theorem}

\usepackage{changebar}
\usepackage{cite}
\ifCLASSINFOpdf
  \usepackage[pdftex]{graphicx}
  \graphicspath{{../pdf/}{../jpeg/}}
  \DeclareGraphicsExtensions{.pdf,.jpeg,.png}
\else
  \usepackage[dvips]{graphicx}
  \graphicspath{{../eps/}}
  \DeclareGraphicsExtensions{.eps}
\fi
\usepackage[cmex10]{amsmath}
\usepackage{array}

\usepackage[tight,footnotesize]{subfigure}
\usepackage{stfloats}
\usepackage{arydshln}
\usepackage{rotating}
\usepackage{microtype}
\def\gap{.37ex}
\abovedisplayskip\gap
\belowdisplayskip\gap
\abovedisplayshortskip\gap
\belowdisplayshortskip\gap

\IEEEoverridecommandlockouts

\usepackage{epstopdf}
\begin{document}
\title{Coding Schemes for a Class of Receiver Message Side Information in AWGN Broadcast Channels}

\author{\IEEEauthorblockN{Behzad Asadi, Lawrence Ong, and Sarah J.\ Johnson}\thanks{This work is supported by the Australian Research Council under grants FT110100195 and DE120100246.}
\IEEEauthorblockA{School of Electrical Engineering and Computer Science, The University of Newcastle, Newcastle, Australia}
Email:{ behzad.asadi@uon.edu.au, lawrence.ong@cantab.net, sarah.johnson@newcastle.edu.au}
}
\maketitle

\begin{abstract}
This paper considers the three-receiver AWGN broadcast channel where the receivers (i) have private-message requests and (ii) know some of the messages requested by other receivers as side information. For this setup, all possible side information configurations have been recently classified into eight groups and the capacity of the channel has been established for six groups (Asadi et al., ISIT 2014). We propose inner and outer bounds for the two remaining groups, groups 4 and 7. A distinguishing feature of these two groups is that the weakest receiver knows the requested message of the strongest receiver as side information while the in-between receiver does not. For group 4, the inner and outer bounds coincide at certain regions. For group 7, the inner and outer bounds coincide, thereby establishing the capacity, for four members out of all eight members of the group; for the remaining four members, the proposed bounds reduce the gap between the best known inner and outer bounds.
\end{abstract}
\vspace{-2pt}
\IEEEpeerreviewmaketitle
\section{Introduction}
We consider \textit{private-message} broadcasting over the three-receiver additive white Gaussian noise broadcast channel (AWGN BC) where the receivers may know some of the source messages a priori. We investigate the capacity of the channel for a class of side information where the weakest receiver knows the requested message of the strongest receiver as side information while the in-between receiver does not.
\vspace{-0pt}
\subsection{Background}
The capacity of BCs \cite{BC} with receiver message side information, where each receiver may know some of the messages requested by other receivers as side information, is of interest due to applications such as multimedia broadcasting with packet loss, and multi-way relay channels \cite{MWRCFullExchange}. The capacity of these channels is known when each receiver needs to decode all the source messages (or equivalently, all the messages not known a priori) \cite{BCwithSI2UsersOechtering,SWoverBC}. Otherwise, the capacity of BCs with receiver message side information is not known in general. 

The capacity of the two-receiver \textit{discrete-memoryless} BC when one of the receivers need not decode all the source messages has been established by Kramer et al. \cite{BCwithSI2UsersKramer}. The capacity of the two-receiver \textit{AWGN} BC is known for all message request and side information configurations \cite{BCwithSI2UsersGeneral}. Oechtering et al. \cite{BCwithSI3UsersCommonMessage} established the capacity of the three-receiver less-noisy and more-capable broadcast channels for some message request and side information configurations where (i) only two receivers possess side information and (ii) the request of the third receiver is only restricted to a common message.
\vspace{-0pt}
\subsection{Existing Results and Contributions}
Considering private-message broadcasting over the three-receiver AWGN BC, Yoo et al. \cite{BCwithSI3UsersPrivateMessage} proposed a separate index and channel coding scheme that achieves within a constant gap of the channel capacity for \textit{all side information configurations}. For this setup, all side information configurations have been recently classified into eight groups and the capacity of the channel has been established for six groups  \cite{Capacity3UsersPrivateMessage}. 

In this paper, we propose new inner and outer bounds for the two remaining groups, groups 4 and 7. For group 4, the proposed inner and outer bounds coincide at certain regions. The inner bound is achieved by two schemes employing dirty paper coding \cite{DPC} with different order of encoding. The outer bound employs the notion of an \textit{enhanced channel} \cite{MIMOBC}, and is shown to be tighter than the best existing one \cite{BCwithSI3UsersPrivateMessage}. For group 7, the proposed inner and outer bounds coincide, thereby establishing the capacity, for four members out of all eight members of the group; for the remaining four members, we improve both the best existing inner bound \cite{Capacity3UsersPrivateMessage} and outer bound~\cite{BCwithSI3UsersPrivateMessage}.

\vspace{-5pt}
\section{System Model}\label{Section:SystemModel}
\begin{figure}[t]
\centering
\includegraphics[width=0.36\textwidth]{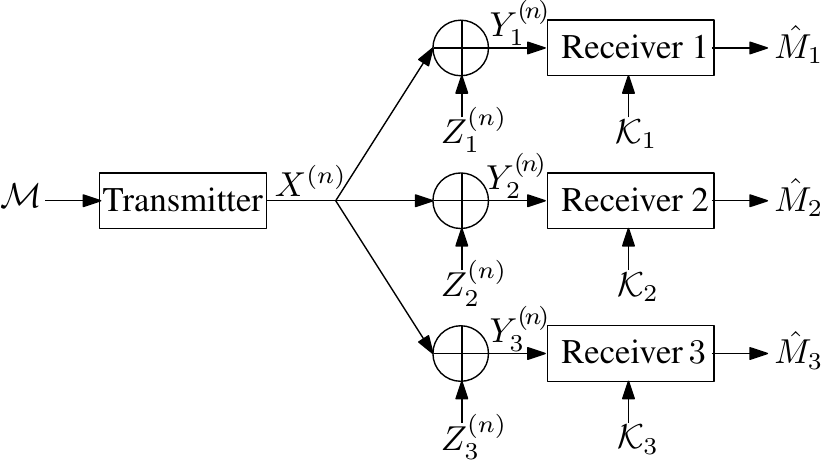}
\vspace{-12pt}
\caption{The three-receiver AWGN broadcast channel with receiver message side information, where $\mathcal{M}=\{M_1,M_2,M_3\}$ is the set of independent messages, each demanded by one receiver, and  $\mathcal{K}_i \subseteq \mathcal{M}\setminus\{M_i\}$ is the set of messages known to receiver $i$ a priori.} 
\vspace{-15pt}
\label{AWGNBCModelFig}
\end{figure}

In the channel model under consideration, as depicted in Fig. \ref{AWGNBCModelFig}, the signals received by receiver $i$, $Y_{i}^{(n)}=\left(Y_{i1},Y_{i2},\ldots,Y_{in}\right)\;i=1,2,3$, is the sum of the codeword transmitted by the sender, $X^{(n)}$, and an i.i.d. noise sequence, $Z_i^{(n)} \;i=1,2,3$, with normal distribution, $Z_i\sim \mathcal{N}\left(0, N_i\right)$. This channel is stochastically degraded, and without loss of generality, we can assume that receiver $1$ is the strongest and receiver $3$ is the weakest in the sense that $N_1\leq N_{2} \leq N_{3}$.

The transmitted codeword has a power constraint of $\sum_{l=1}^{n}E\left(X_l^2\right)\hspace{-3pt}\leq\hspace{-3pt}nP$ and is a function of source messages $\mathcal{M}=\{M_1,M_2,M_3\}$. The messages are independent, and $M_i$ is intended for receiver $i$ at rate $R_i$ bits per channel use, i.e., $m_i\in\{1,2,\ldots,2^{nR_i}\}$. The \textit{capacity} of the channel is the closure of the set of all rate triples $(R_1,R_2,R_3)$ that are \textit{achievable} in the Shannon sense \cite{BC}.

We define the \textit{knows} set $\mathcal{K}_i$ as the set of messages known to receiver $i$. The side information configuration of the channel is modeled by a side information graph, $\mathcal{G}=\left(\mathcal{V}_\mathcal{G},\mathcal{A}_\mathcal{G}\right)$, where $\mathcal{V}_\mathcal{G}$ is the set of \textit{vertices} and $\mathcal{A}_\mathcal{G}$ is the set of \textit{arcs}. Vertex $i$ represents both $M_i$ and receiver $i$ requesting it. An arc from vertex $i$ to vertex $j$, denoted by $(i\rightarrow j)$, exists if and only if receiver $i$ knows $M_j$. The set of out-neighbors of vertex $i$ is then $\mathcal{O}_i\triangleq\{j\mid (i\rightarrow j)\in\mathcal{A}_\mathcal{G}\}=\{j\mid M_j\in\mathcal{K}_i\}$. For instance, in the following side information graph 
\vspace{-5pt}
\begin{center}
\includegraphics[width=0.135\textwidth]{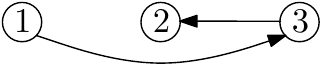}
\end{center}
\vspace{-7pt}
receiver 1 knows $M_3$, and receiver 3 knows $M_2$.

\begin{figure}[t]
\centering
\vspace{5pt}
\includegraphics[width=0.315\textwidth]{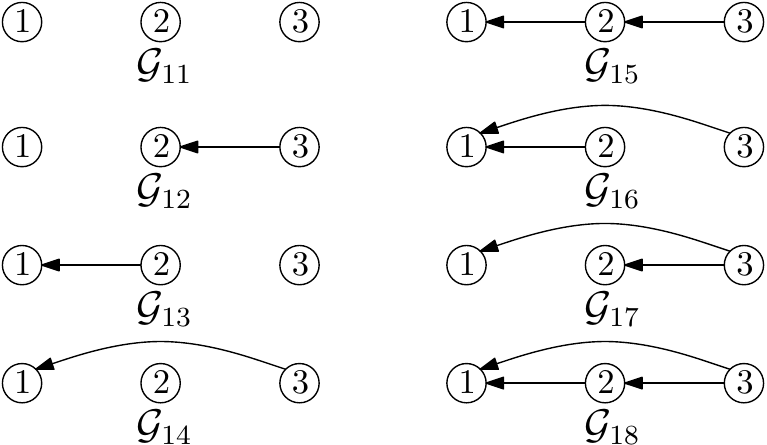}
\vspace{-8pt}
\caption{Group leaders, capturing if each receiver knows the message(s) requested by stronger receiver(s).}
\label{Fig:GroupLeaders}
\vspace{-12pt}
\end{figure}

\begin{figure}[t]
\centering
\includegraphics[width=0.315\textwidth]{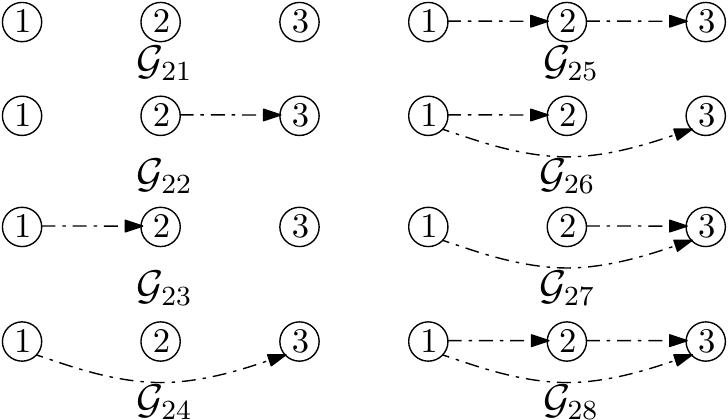}
\vspace{-11pt}
\caption{Graphs, capturing if each receiver knows the message(s) requested by weaker receiver(s).} 
\label{Fig:Subgraphs}
\vspace{-15pt}
\end{figure}

All possible side information graphs for the three-receiver case, i.e., $\mathcal{V}_\mathcal{G}=\{1,2,3\}$ have been classified into eight groups \cite{Capacity3UsersPrivateMessage}. Any side information graph is the union of $\mathcal{G}_{1j}$ (depicted in Fig. \ref{Fig:GroupLeaders}) and $\mathcal{G}_{2k}$ (depicted in Fig. \ref{Fig:Subgraphs}) for some unique $j$ and $k$ where $j,k\in\{1,2,\ldots,8\}$. According to this classification, the side information graphs $\{\mathcal{G}_{1j}\}_{j=1}^{8}$  are considered as the group leaders, and group $j$ consists of the side information graphs formed by the union of $\mathcal{G}_{1j}$ with each of the $\{\mathcal{G}_{2k}\}_{k=1}^8$. In this work, we investigate the capacity of the channel for group 4, $\{\mathcal{G}_{14}\cup\mathcal{G}_{2k}\}_{k=1}^8$, and group 7, $\{\mathcal{G}_{17}\cup\mathcal{G}_{2k}\}_{k=1}^8$.

\begin{table*}[t]
\begin{footnotesize}
\caption{Group 4: Proposed transmission schemes and inner bounds}
\vspace{-14pt}
\begin{center}
{\renewcommand{\arraystretch}{2}
\begin{tabular}{|c|l|l|}
\cline{2-3}
\multicolumn{1}{c|}{}&\hspace{40pt}{Transmission Scheme\,1 \& Inner Bound\,1 ($\mathscr{R}_{\text{in}_1}$)} &{\hspace{35pt}Transmission Scheme\,2 \& Inner Bound\,2 ($\mathscr{R}_{\text{in}_2}$)}\\
\cline{2-3}
\multicolumn{1}{c|}{} &{$\color{blue}x_1^{(\hspace{-1pt}n\hspace{-1pt})}\hspace{-2pt}\left([m_1,m_3]\right)+x_2^{(\hspace{-1pt}n\hspace{-1pt})}\hspace{-2pt}\left(m_2,x_1^{(\hspace{-1pt}n\hspace{-1pt})}\hspace{-2pt}\left([m_1,m_3]\right)\right)$}&{\color{blue}{$x_1^{(\hspace{-1pt}n\hspace{-1pt})}\hspace{-2pt}\left([m_1,m_3],x_2^{(\hspace{-1pt}n\hspace{-1pt})}\hspace{-2pt}\left(m_2\right)\right)+x_2^{(\hspace{-1pt}n\hspace{-1pt})}\hspace{-2pt}\left(m_2\right)$}}\\

\multicolumn{1}{c|}{}&
{$\color{blue}\hspace{-18pt}\underset{\hspace{15pt}i\in\{1,3\}\setminus\mathcal{O}_1}{\sum}\hspace{-15pt}R_i{<}C\hspace{-2pt}\left(\hspace{-1pt}\frac{(1-\alpha)P}{\alpha P+N_1}\hspace{-1pt}\right)\hspace{-2pt},
R_2{<}C\hspace{-2pt}\left(\hspace{-1pt}\frac{\alpha P}{N_2}\hspace{-1pt}\right)\hspace{-2pt}, 
R_3{<}C\hspace{-2pt}\left(\frac{(1-\alpha)P}{\alpha P+N_3}\right)$}&{\color{blue}{$\hspace{-18pt}\underset{\hspace{15pt}i\in\{1,3\}\setminus\mathcal{O}_1}{\sum}\hspace{-15pt}R_i {<}C\hspace{-2pt}\left(\hspace{-1pt}\frac{\gamma P}{N_1}\hspace{-1pt}\right)\hspace{-2pt},
R_2{<}C\hspace{-2pt}\left(\hspace{-1pt}\frac{(1-\gamma)P}{\gamma P+N_2}\hspace{-1pt}\right)\hspace{-2pt}, 
R_3{<}C\hspace{-2pt}\left(\frac{\gamma P}{N_3}\right)$}}\\
\hline
$\hspace{-2pt}\mathcal{G}_{14}\hspace{-1pt}\cup\hspace{-1pt}\mathcal{G}_{22}\hspace{-5pt}$&$\color{blue}x_1^{(\hspace{-1pt}n\hspace{-1pt})}\hspace{-2pt}\left([m_1\hspace{-2pt}\oplus\hspace{-2pt} m_{31},m_{30}]\right)+x_2^{(\hspace{-1pt}n\hspace{-1pt})}\hspace{-2pt}\left([m_2,m_{31}],x_1^{(\hspace{-1pt}n\hspace{-1pt})}\hspace{-2pt}\left([m_1\hspace{-2pt}\oplus\hspace{-2pt} m_{31}, m_{30}]\right)\right)\hspace{-5pt}$&\color{blue}{$x_1^{(\hspace{-1pt}n\hspace{-1pt})}\hspace{-2pt}\left([m_1\hspace{-2pt}\oplus\hspace{-2pt} m_{31},m_{30}],x_2^{(\hspace{-1pt}n\hspace{-1pt})}\hspace{-2pt}\left([m_2,m_{31}]\right)\right)+x_2^{(\hspace{-1pt}n\hspace{-1pt})}\hspace{-2pt}\left([m_2,m_{31}]\right)$}\\
\cline{1-1}
$\hspace{-2pt}\mathcal{G}_{14}\hspace{-1pt}\cup\hspace{-1pt}\mathcal{G}_{25}\hspace{-5pt}$&$\color{blue}R_1\hspace{-3pt}<\hspace{-2pt}C\hspace{-2pt}\left(\hspace{-1pt}\frac{(1-\alpha)P}{\alpha P+N_1}\hspace{-1pt}\right)\hspace{-2pt},\hspace{-2pt}\underset{i\notin\mathcal{O}_1}{\sum}\hspace{-2pt}R_i{<}C\hspace{-2pt}\left(\hspace{-1pt}\frac{P}{N_1}\hspace{-1pt}\right)\hspace{-2pt},
R_2\hspace{-3pt}<\hspace{-2pt}C\hspace{-2pt}\left(\hspace{-1pt}\frac{\alpha P}{N_2}\hspace{-1pt}\right)\hspace{-2pt},
R_3\hspace{-2pt}<\hspace{-2pt}C\hspace{-2pt}\left(\frac{(1-\alpha)P}{\alpha P+N_3}\right)\hspace{-5pt}$&\color{blue}{$R_1\hspace{-2pt}<\hspace{-3pt}C\hspace{-2pt}\left(\hspace{-1pt}\frac{\gamma P}{N_1}\hspace{-1pt}\right)\hspace{-2pt}, 
\hspace{-2pt}\underset{i\notin\mathcal{O}_1}{\sum}\hspace{-2pt}R_i{<}C\hspace{-2pt}\left(\hspace{-1pt}\frac{P}{N_1}\hspace{-1pt}\right)\hspace{-2pt},
R_2\hspace{-3pt}<\hspace{-3pt}C\hspace{-2.5pt}\left(\hspace{-1pt}\frac{(1-\gamma)P}{\gamma P+N_2}\hspace{-1pt}\right)\hspace{-2pt},
R_3\hspace{-3pt}<\hspace{-3pt}C\hspace{-2pt}\left(\frac{\gamma P}{N_3}\right)\hspace{-5pt}$}\\
\hline
\end{tabular}}
\label{Table:group4schemes}
\end{center}
\end{footnotesize}
\vspace{-21pt}
\end{table*}
\vspace{-0pt}
\section{Group 4: Proposed Inner and Outer Bounds}\label{Section:Group4}
In this section, we first propose an inner bound and an outer bound for group 4. We next characterize the regions where these bounds coincide.
\vspace{-0pt}
\subsection{Proposed Inner Bound}
The proposed inner bound for group 4, stated as Theorem~\ref{theorem:in1}, is the convex hull of the union of two regions achieved by the transmission schemes presented in Table~\ref{Table:group4schemes}. These schemes are constructed using rate splitting, index coding \cite{IndexCoding}, multiplexing coding \cite{MultiplexedCoding}, dirty paper coding \cite{DPC}, and superposition coding. In rate splitting, the message $M_i$ is divided into a set of independent messages $\{M_{il}\}_{l=0}^{L}$ with rates $\{R_{il}\}_{l=0}^{L}$ such that $R_i=\sum_{l=0}^{L}R_{il}$. In index coding, the transmitter XORs the messages to accomplish compression prior to channel coding; in Table~\ref{Table:group4schemes}, $\oplus$ denotes the bitwise XOR with zero padding for messages of unequal length. In multiplexing coding, multiple messages are first bijectively mapped to a single message and a codebook is then constructed for this message; in Table~\ref{Table:group4schemes}, square brackets, $[\cdot]$, denote a bijective map. Dirty paper coding is used when the channel between a transmitter and a receiver undergoes an interference $s^{(n)}$ which is known non-causally at the transmitter; codewords using this scheme are denoted by $x_i^{(n)}\hspace{-2pt}\left(m,s^{(n)}\right)$ where $m$ is the transmitted message.

The codebook of transmission scheme 1 is formed from the linear superposition of two subcodebooks. The first subcodebook consists of i.i.d.\ codewords $x_1^{(n)}$ generated according to $X_1\hspace{-2pt}\sim\hspace{-2pt}\mathcal{N}\left(0,(1\hspace{-2pt}-\hspace{-2pt}\alpha)P\right)$ where $0\hspace{-2pt}\leq\hspace{-2pt}\alpha\hspace{-2pt}\leq\hspace{-2pt}1$. By treating $x_1^{(n)}$ as interference for receiver 2, which is known non-causally at the transmitter, the second subcodebook of this scheme is constructed using dirty paper coding. The \textit{auxiliary random variable} in dirty paper coding is defined as $U\hspace{-2pt}=\hspace{-2pt}X_2+\beta X_1$ where  $X_2\hspace{-2pt}\sim\hspace{-2pt}\mathcal{N}\left(0,\alpha P\right)$ is independent of $X_1$, and $\beta=\frac{\alpha P}{\alpha P+N_2}$. 

The codebook of transmission scheme 2 is also formed from the linear superposition of two subcodebooks. The second subcodebook is formed from i.i.d.\ codewords $x_2^{(n)}$ generated according to $X_2\hspace{-2pt}\sim\hspace{-2pt}\mathcal{N}\left(0, (1\hspace{-2pt}-\hspace{-2pt}\gamma)P\right)$ where $0\hspace{-2pt}\leq\hspace{-2pt}\gamma\hspace{-2pt}\leq\hspace{-2pt}1$. By treating $x_2^{(n)}$ as interference for receiver 3, which is known non-causally at the transmitter, the first subcodebook of this scheme is constructed using dirty paper coding. The auxiliary random variable in dirty paper coding is defined as $U=X_1+\beta X_2$ where $X_1\sim\mathcal{N}\left(0,\gamma P\right)$ is independent of $X_2$, and $\beta=\frac{\gamma P}{\gamma P+N_3}$.


There are two members in this group, $\mathcal{G}_{14}\cup\mathcal{G}_{22}$ and $\mathcal{G}_{14}\cup\mathcal{G}_{25}$, that use modified versions of the schemes. In these modified schemes, using rate splitting, the message $M_3$ is divided into $M_{30}$ and $M_{31}$.

As it is seen, the two transmission schemes for each member employ dirty paper coding with different order of encoding. These two schemes can be combined using the approach shown by Oechtering et al. \cite{OechteringG14G24}.

Here, we state the proposed inner bound for group 4.
\begin{theorem}\label{theorem:in1}
The rate triple $(R_1,R_2,R_3)$ for a member of group 4 is achievable, if it lies in the convex hull of the union of inner bound 1, $\mathscr{R}_{\text{in}_1}$, and inner bound 2, $\mathscr{R}_{\text{in}_2}$ (i.e., co($\mathscr{R}_{\text{in}_1}\hspace{-2pt}\cup\hspace{-2pt}\mathscr{R}_{\text{in}_2}$)), given in Table~\ref{Table:group4schemes}. For each member, $\mathscr{R}_{\text{in}_1}$ is the set of all rate triples, each satisfying the conditions in the first column of the respective row for some $0\leq\alpha\leq1$, and $\mathscr{R}_{\text{in}_2}$ is the set of all rate triples, each satisfying the conditions in the second column of the respective row for some $0\leq\gamma\leq1$, where $C(q) \hspace{-3pt}\triangleq\hspace{-3pt}\frac{1}{2}\log(1\hspace{-2pt}+\hspace{-2pt}q)$.
\end{theorem}
\begin{IEEEproof}
The achievability of $\mathscr{R}_{\text{in}_1}$ is verified by using the following decoding methods. At receiver 1, $x_1^{(n)}$ is decoded while $x^{(n)}_2$ is treated as noise, and then $x^{(n)}_2$ is decoded. At receiver 2, $m_2$ is decoded without being affected by $x_1^{(n)}$ due to dirty paper coding. At receiver 3, $x_1^{(n)}$ is just decoded while $x^{(n)}_2$ is treated as noise. 

The achievability of $\mathscr{R}_{\text{in}_2}$ is verified by using the following decoding methods. At receiver 3, $m_3$ is decoded without being affected by $x_2^{(n)}$ due to dirty paper coding. At receiver 2, $x_2^{(n)}$ is decoded while $x_1^{(n)}$ is treated as noise. At receiver 1, $x_2^{(n)}$ is first decoded while $x_1^{(n)}$ is treated as noise, and then $x_1^{(n)}$ is decoded. 

Note that the receivers utilize their side information during channel coding. Also, for $\mathcal{G}_{14}\hspace{-2pt}\cup\hspace{-2pt}\mathcal{G}_{22}$ and $\mathcal{G}_{14}\hspace{-2pt}\cup\hspace{-2pt}\mathcal{G}_{25}$, Fourier-Motzkin elimination is used subsequent to channel decoding to obtain $\mathscr{R}_{\text{in}_1}$ and $\mathscr{R}_{\text{in}_2}$ in terms of $(R_1,R_2,R_3)$.
\end{IEEEproof}

In this paper, all rate bound derivations use standard techniques, and are omitted due to space limitations.

\begin{table*}[t]
\begin{footnotesize}
\caption{Group 7: Proposed transmission schemes and inner bounds}
\vspace{-14pt}
\begin{center}
{\renewcommand{\arraystretch}{1.33}
\begin{tabular}{|l|l|l|l|}
\hline
Member&Graph&Transmission Scheme&Inner Bound ($\mathscr{R}'_{\text{in}}$)\\
\hline
$\mathcal{G}_{17}\cup\mathcal{G}_{21}$&\hspace{-5pt}\raisebox{-0.5ex}{\includegraphics[width=0.075\textwidth]{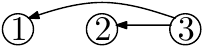}}$\hskip-5pt$&\multirow{2}{*}{$\color{blue}x_1^{(n)}\left([m_{10},m_{30}]\right)+x_2^{(n)}\left([m_2, m_{11},m_{31}]\right)$}&\multirow{3}{*}{\hspace{-4pt}\color{blue}{$R_2+\hspace{-15pt}\underset{\hspace{15pt}i\in\{1,3\}\setminus\mathcal{O}_1}{\sum}\hspace{-15pt}R_i \overset{{\color{black}(a)}}{<}C\left(\frac{\left(1-\alpha\right)P}{\alpha P+N_2}\right)+C\left(\frac{\alpha P}{N_1}\right), R_2\overset{{\color{black}(b)}}{<} C\left(\frac{\left(1-\alpha\right)P}{\alpha P+N_2}\right),
$}}\\
\cline{1-2}
$\mathcal{G}_{17}\cup\mathcal{G}_{23}$&\hspace{-5pt}\raisebox{-0.5ex}{\includegraphics[width=0.075\textwidth]{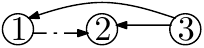}}$\hskip-5pt$&&\\
\cline{1-3}
$\mathcal{G}_{17}\cup\mathcal{G}_{24}$&\hspace{-5pt}\raisebox{-1ex}{\includegraphics[width=0.075\textwidth]{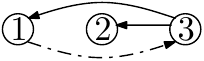}}$\hskip-5pt$&\multirow{2}{*}{$\color{blue}x_1^{(n)}\left([m_{10},m_{30}]\right)+x_2^{(n)}\left([m_2, m_{11}\hspace{-3pt}\oplus\hspace{-2pt}m_{31}]\right)$}&\multirow{2}{*}{\hspace{-4pt}\color{blue}{$R_2+R_3\overset{{\color{black}(c)}}{<}C\left(\frac{\left(1-\alpha\right)P}{\alpha P+N_2}\right)+C\left(\frac{\alpha P}{N_3}\right), R_3\overset{{\color{black}(d)}}{<}C\left(\frac{P}{N_3}\right)$}}\\
\cline{1-2}
$\mathcal{G}_{17}\cup\mathcal{G}_{26}$&\hspace{-5pt}\raisebox{-1ex}{\includegraphics[width=0.075\textwidth]{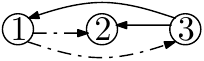}}$\hskip-5pt$&&\\
\hline
$\mathcal{G}_{17}\cup\mathcal{G}_{22}$&\hspace{-5pt}\raisebox{-0.5ex}{\includegraphics[width=0.075\textwidth]{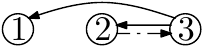}}$\hskip-5pt$&\multirow{4}{*}{$\color{blue}x_1^{(n)}\left([m_1,m_2\oplus m_{3}]\right)+x_2^{(n)}\left(m_2\oplus m_{3}\right)$}&\multirow{3}{*}{\hspace{-4pt}\color{blue}{$R_1< C\left(\frac{\alpha P}{N_1}\right), \hspace{-15pt}\underset{\hskip15pt i\in\{1,3\}\setminus\mathcal{O}_1}{\sum}\hspace{-15pt}R_i< C\left(\frac{P}{N_1}\right),$}}\\
\cline{1-2}
$\mathcal{G}_{17}\cup\mathcal{G}_{25}$&\hspace{-5pt}\raisebox{-0.5ex}{\includegraphics[width=0.075\textwidth]{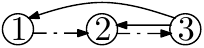}}$\hskip-5pt$&&\\
\cline{1-2}
$\mathcal{G}_{17}\cup\mathcal{G}_{27}$&\hspace{-5pt}\raisebox{-1ex}{\includegraphics[width=0.075\textwidth]{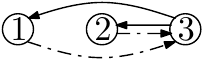}}$\hskip-5pt$&&\multirow{2}{*}{\hspace{-4pt}\color{blue}{$R_2< C\left(\frac{\left(1-\alpha\right)P}{\alpha P+N_2}\right),
R_3< C\left(\frac{P}{N_3}\right)\hspace{-5pt}$}}\\
\cline{1-2}
$\mathcal{G}_{17}\cup\mathcal{G}_{28}$&\hspace{-5pt}\raisebox{-1ex}{\includegraphics[width=0.075\textwidth]{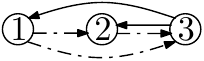}}$\hskip-5pt$&&\\
\hline
\end{tabular}}
\label{group7schemes}
\end{center}
\end{footnotesize}
\vspace{-20pt}
\end{table*}
\vspace{-0pt}
\subsection{Proposed Outer Bound}
The proposed outer bound for group 4, stated as Theorem~\ref{theorem:out1}, is formed from the intersection of two outer bounds.
\begin{theorem}\label{theorem:out1} 
If the rate triple $(R_1,R_2,R_3)$ is achievable for a member of group 4, then it must lie in $\mathscr{R}_{\text{out}_1}\cap\mathscr{R}_{\text{out}_2}$ where $\mathscr{R}_{\text{out}_1}$ is the set of all rate triples, each satisfying
\begin{align}
&R_1\leq C\left(\frac{P}{N_1}\right),\label{outer11}\\
&R_2\leq C\left(\frac{\alpha P}{N_2}\right),\label{outer12}\\
&R_3\leq C\left(\frac{(1-\alpha)P}{\alpha P+N_3}\right),\label{outer13}
\end{align}
for some $0\leq\alpha\leq1$, and $\mathscr{R}_{\text{out}_2}$ is the capacity of the enhanced channel for the member obtained by decreasing the received noise variance of receiver 3 from $N_3$ to $N_2$.
\end {theorem}
\begin{IEEEproof}
Condition \eqref{outer11} in $\mathscr{R}_{\text{out}_1}$ is due to the point-to-point channel capacity between the transmitter and receiver 1. Conditions \eqref{outer12} and \eqref{outer13} follow from the capacity of the two-receiver AWGN BC (from the transmitter to receivers 2 and 3) where only the stronger receiver (receiver 2) may know the requested message of the weaker receiver (receiver 3) as side information. In this group, the side information of receivers 2 and 3 about each other's requested messages has this property.

Outer bound 2, $\mathscr{R}_{\text{out}_2}$, is developed using the idea of enhanced channel \cite{MIMOBC}. The capacity of the enhanced channel is an outer bound to the capacity of the original channel. Since the received noise variance of the two weakest receivers in the defined enhanced channel are equal, this channel can also be considered as a member of group 5 or 3 depending on whether receiver 2 in the original channel knows $M_3$ or not, respectively. The capacity of the channel with a side information configuration in group 3 or 5 is known \cite{Capacity3UsersPrivateMessage}. For example, the enhanced channel for $\mathcal{G}_{14}\cup\mathcal{G}_{21}$ can be considered as $\mathcal{G}_{13}\cup\mathcal{G}_{21}$, and the one for $\mathcal{G}_{14}\cup\mathcal{G}_{22}$ as $\mathcal{G}_{15}\cup\mathcal{G}_{21}$. 
\end{IEEEproof}
\vspace{-0pt}
\subsection{Evaluation of the Proposed Inner and Outer Bounds} 
In this subsection, we first show that the proposed outer bound is tighter than the best existing one. We next characterize the regions where the proposed inner and outer bounds coincide.

In order to prove that our proposed outer bound is tighter than the best existing one, we show that, for any condition that must be met in the best existing outer bound, the proposed outer bound includes some more restrictive conditions. We present the proof for $\mathcal{G}_{14}\hspace{-2pt}\cup\hspace{-2pt}\mathcal{G}_{21}$ in the following; the proof for the other members is similar. Our proposed outer bound is the intersection of the bound given in \eqref{outer11}--\eqref{outer13} and the capacity of the enhanced channel for $\mathcal{G}_{14}\cup\mathcal{G}_{21}$. According to the enhanced channel, if the rate triple $(R_1,R_2,R_3)$ is achievable for $\mathcal{G}_{14}\cup\mathcal{G}_{21}$, it must satisfy
\vspace{-3pt}
\begin{align}
R_1+R_3&\leq C\left(\frac{\gamma P}{N_1}\right),\label{outer14}\\
R_2&\leq C\left(\frac{(1-\gamma)P}{\gamma P+N_2}\right),\label{outer15}\\
R_3&\leq C\left(\frac{\gamma P}{N_2}\right),\label{outer16}
\end{align}
for some $0\leq\gamma\leq1$.
Based on the best existing outer bound \cite{BCwithSI3UsersPrivateMessage} if the rate triple $(R_1,R_2,R_3)$ is achievable, it must satisfy
\begin{equation}\label{bestpriorouter}
\sum_{i\in\mathcal{V}_\mathcal{S}}\hspace{-2pt}{R_i}\hspace{-2pt}\leq\hspace{-2pt}\underset{i\in\mathcal{V}_\mathcal{S}}{\max}\,C\left(\frac{P}{N_i}\right),
\end{equation}
for all induced acyclic subgraphs, $\mathcal{S}$, of the side information graph. Then, for $\mathcal{G}_{14}\hspace{-2pt}\cup\hspace{-2pt}\mathcal{G}_{21}$, if the rate triple $(R_1,R_2,R_3)$ is achievable, it must satisfy $R_3\leq C\left(P/N_3\right)$, $R_2+R_3\leq C(P/N_2)$, and $R_1+R_2+R_3\leq C(P/N_1)$. Concerning $R_3\leq C\left(P/N_3\right)$, if condition \eqref{outer13} in $\mathscr{R}_{\text{out}_1}$ is satisfied, this condition is also satisfied.  Conditions \eqref{outer12} and \eqref{outer13} in $\mathscr{R}_{\text{out}_1}$ are more restrictive than $R_2+R_3\leq C(P/N_2)$, and conditions \eqref{outer14} and \eqref{outer15} in $\mathscr{R}_{\text{out}_2}$ are more restrictive than $R_1+R_2+R_3\leq C(P/N_1)$. This completes the proof for $\mathcal{G}_{14}\cup\mathcal{G}_{21}$.

\begin{figure}[t]
\centering
\vspace{-3pt}
\hspace{-0pt}
\includegraphics[width=0.2\textwidth]{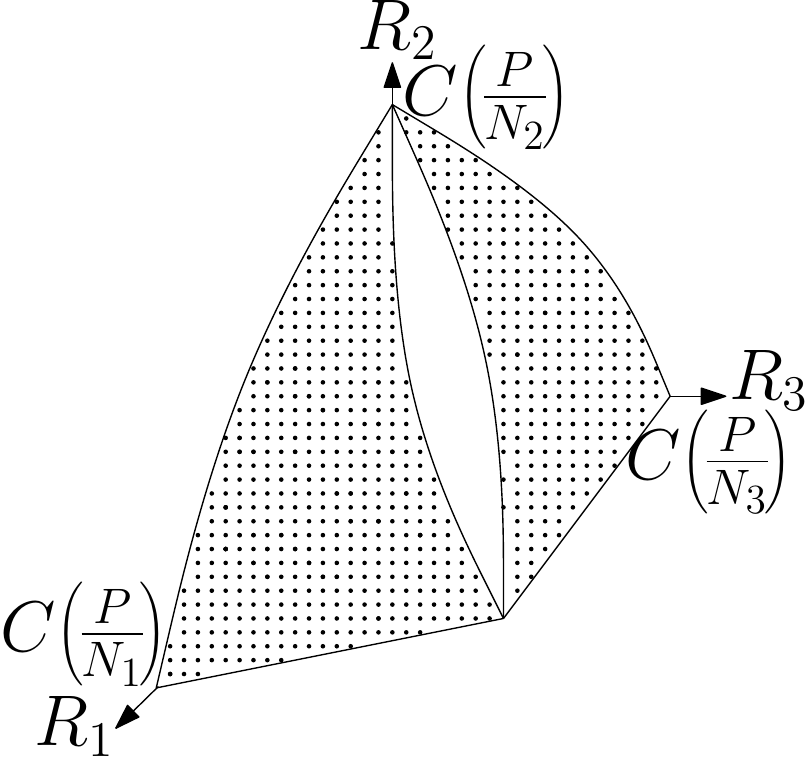}
\vspace{-12pt}
\caption{The proposed bounds are tight at the dotted regions for group 4.} 
\vspace{-15pt}
\label{threedimension}
\end{figure}

Here, we characterize the certain regions where the proposed inner and outer bounds coincide. For any fixed $R_1$ where $0\leq\hspace{-2pt}R_1\hspace{-2pt}\leq C\hspace{-0pt}(\frac{P}{N_1})$, the proposed bounds are tight when $R_3\hspace{-2pt}\leq\hspace{-2pt}R_{\text{thr}_3}$ or $R_3\hspace{-2pt}\geq\hspace{-2pt}R'_{\text{thr}_3}$ where $R_{\text{thr}_3}\hspace{-2pt}\leq\hspace{-2pt}R'_{\text{thr}_3}$; or similarly, when $R_2\leq R_{\text{thr}_2}$ or $R_2\geq R'_{\text{thr}_2}$ where $R_{\text{thr}_2}\leq R'_{\text{thr}_2}$. The thresholds are functions of $R_1$. The same behavior can be observed for any fixed $R_i\;i\in\{1,2,3\}$ on the $R_j\hspace{-2pt}-\hspace{-2pt}R_k$ plane $j,k\hspace{-2pt}\in\hspace{-2pt}\{1,2,3\}\hspace{-2pt}\setminus\hspace{-2pt}\{i\}$. Fig. \ref{threedimension} illustrates this behavior for group 4.

\begin{figure}[b]
\vspace{-15pt}
\hspace{-13pt}
\includegraphics[width=0.55\textwidth]{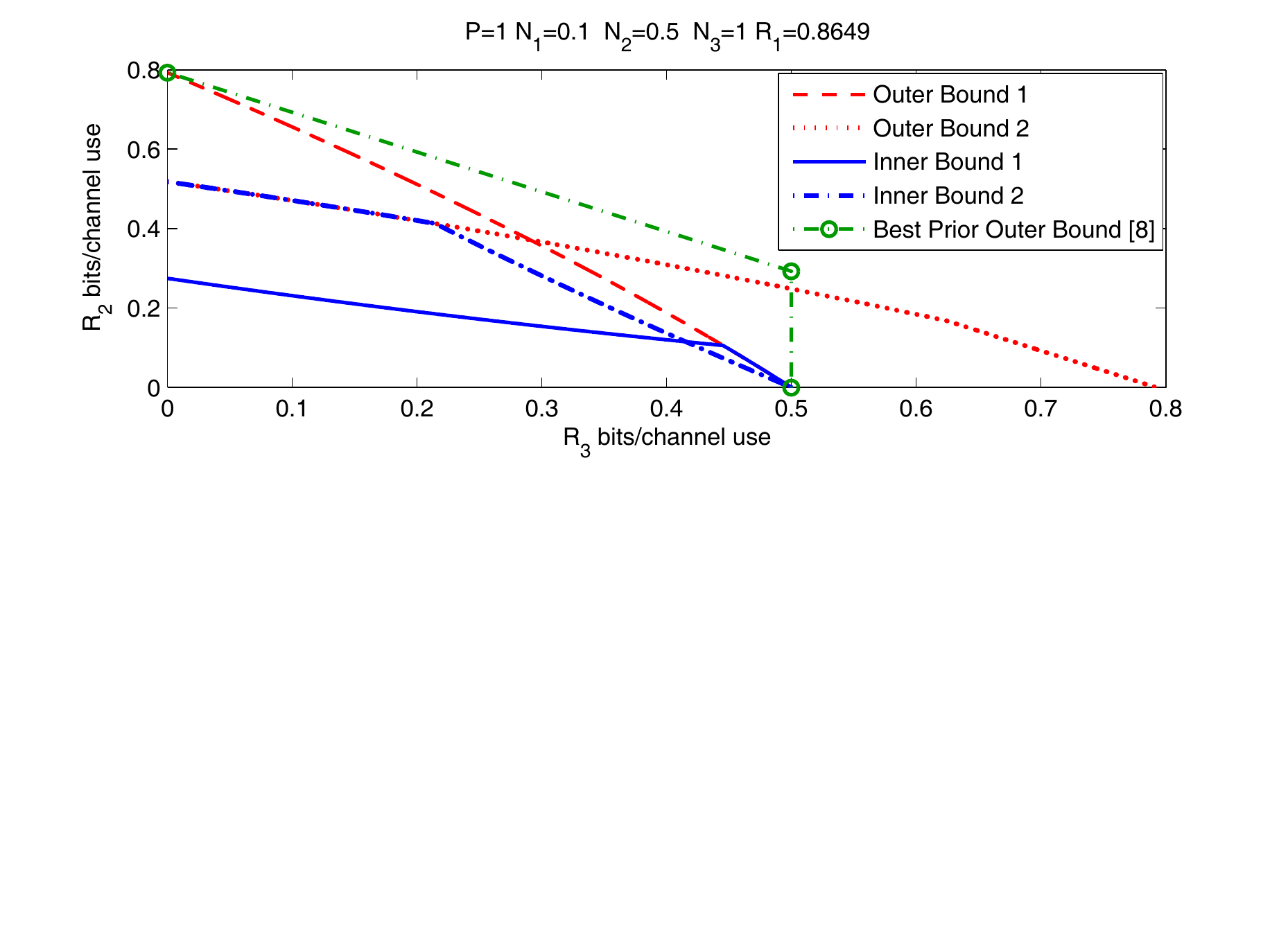}
\vspace{-125pt}
\caption{Inner and outer bounds for $\mathcal{G}_{14}\cup\mathcal{G}_{21}$. The proposed inner bound is the convex hull of the union of inner bounds 1 and 2, and the proposed outer bound is the intersection of outer bounds 1 and 2.} 
\label{Group4comparison}
\end{figure}

Due to space limitations, we present only the thresholds on $R_3$ for $\mathcal{G}_{14}\cup\mathcal{G}_{21}$ as an example. For $R_1=0$, $R_{\text{thr}_3}\hspace{-3pt}=\hspace{-3pt}R'_{\text{thr}_3}=0$, and $\mathscr{R}_{\text{in}_1}$ and $\mathscr{R}_{\text{out}_1}$ coincide. For $0\hspace{-2pt}<\hspace{-2pt}R_1\hspace{-2pt}<\hspace{-2pt}C(\frac{P}{N_1})\hspace{-2pt}-\hspace{-2pt}C(\frac{P}{N_3})$, $R_{\text{thr}_3}\hspace{-3pt}=\hspace{-3pt}C(\frac{\gamma^{\star}P}{N_3})$ where $\gamma^{\star}$ satisfies $R_1\hspace{-3pt}=\hspace{-3pt}C(\frac{\gamma^{\star}P}{N_1})\hspace{-3pt}-\hspace{-3pt}C(\frac{\gamma^{\star}P}{N_3})$, and $R'_{\text{thr}_3}\hspace{-3pt}=\hspace{-3pt}C(\frac{(1-\alpha^{\star})P}{\alpha^{\star}P+N_3})$ where $\alpha^{\star}$ satisfies $R_1\hspace{-3pt}=\hspace{-3pt}C(\frac{(1-\alpha^{\star})P}{\alpha^{\star}P+N_1})-C(\frac{(1-\alpha^{\star})P}{\alpha^{\star}P+N_3})$. For $C(\frac{P}{N_1})\hspace{-2pt}-\hspace{-2pt}C(\frac{P}{N_3})\hspace{-2pt}\leq\hspace{-2pt}R_1\hspace{-2pt}<\hspace{-2pt}C(\frac{P}{N_1})$, we have $R_{\text{thr}_3}\hspace{-3pt}=\hspace{-3pt}R'_{\text{thr}_3}=C(\frac{P}{N_1})\hspace{-3pt}-\hspace{-3pt}R_1$, and $\mathscr{R}_{\text{in}_2}$ and $\mathscr{R}_{\text{out}_2}$ coincide. For $\mathcal{G}_{14}\cup\mathcal{G}_{21}$, Fig. \ref{Group4comparison} shows that the proposed outer bound is strictly tighter than best existing one. This figure also shows that for a fixed $0\hspace{-2pt}<\hspace{-2pt}R_1\hspace{-2pt}<\hspace{-2pt}C(\frac{P}{N_1})\hspace{-2pt}-\hspace{-2pt}C(\frac{P}{N_3})$, the proposed bounds coincide when $R_2$ or $R_3$ is below or above certain thresholds.
\vspace{-3pt}
\section{Group 7: Proposed Inner and Outer Bounds}\label{Section:Group7}
In this section, we propose an inner bound and an outer bound for group 7, and compare them with the best prior ones.
\vspace{-16pt}
\subsection{Proposed Inner Bound}
The proposed inner bound for group 7, stated as Theorem \ref{theorem:in2}, is achieved by the proposed transmission schemes presented in Table~\ref{group7schemes}. These schemes are constructed using rate splitting, index coding, multiplexing coding and superposition coding. Each transmission scheme includes two subcodebooks; the first subcodebook consists of i.i.d. codewords generated according to $X_1\sim\mathcal{N}(0,\alpha P)$, and the second subcodebook consists of i.i.d. codewords generated independently according to $X_2\sim\mathcal{N}(0,(1-\alpha)P)$ where $0\leq\alpha\leq1$. 
\begin{theorem}\label{theorem:in2} 
The rate triple $(R_1,R_2,R_3)$ for a member of group 7 is achievable, if it lies in $\mathscr{R}'_{\text{in}}$, given in Table~\ref{group7schemes}, where $\mathscr{R}'_{\text{in}}$ is the set of all rate triples, each satisfying the conditions in the respective row of Table~\ref{group7schemes} for some $0\leq\alpha\leq1$.
\end{theorem}
\begin{IEEEproof}
The achievability of $\mathscr{R}'_{\text{in}}$ for the members using rate splitting is verified by employing successive decoding followed by Fourier-Motzkin elimination. For these members, at receivers 1 and 3, $x_2^{(n)}$ is first decoded while $x_1^{(n)}$ is treated as noise, and then $x_1^{(n)}$ is decoded. At receiver 2, $x_2^{(n)}$ is just decoded while $x_1^{(n)}$ is treated as noise. The achievability of $\mathscr{R}'_{\text{in}}$ for the members not using rate splitting is verified by employing simultaneous decoding \cite[p.\ 88]{NITBook} at receivers 1 and 3, and successive decoding at receiver 2 where $x_2^{(n)}$ is decoded while $x_1^{(n)}$ is treated as noise. Note that the receivers utilize their side information during channel decoding.
\end{IEEEproof}

\subsection{Proposed Outer Bound}
The proposed outer bound for group 7, stated as Theorem~\ref{theorem:out2}, is formed from the intersection of two outer bounds. One of them is the best existing outer bound given in \eqref{bestpriorouter}. 
\begin{theorem}\label{theorem:out2} 
If the rate triple $(R_1,R_2,R_3)$ is achievable for a members of group 7, then it must lie in $\mathscr{R}'_{\text{out}_1}\cap\mathscr{R}'_{\text{out}_2}$ where $\mathscr{R}'_{\text{out}_1}$ is the set of all rate triples, each satisfying
\begin{align}
&R_1\leq C\left(\frac{\alpha P}{N_1}\right),\label{outer21}\\
&R_2\leq C\left(\frac{\left(1-\alpha\right)P}{\alpha P+N_2}\right),\label{outer22}\\
&R_3\leq C\left(\frac{P}{N_3}\right),\label{outer23}
\end{align}
for some $0\leq\hspace{-2pt}\alpha\hspace{-2pt}\leq1$, and $\mathscr{R}'_{\text{out}_2}$ is the outer bound given in \eqref{bestpriorouter}.
\end{theorem}
\begin{IEEEproof}
Conditions \eqref{outer21} and \eqref{outer22} follow from the capacity of the two-receiver AWGN BC (from the transmitter to receivers 1 and 2) where only the stronger receiver (receiver 1) may know the requested message of the weaker receiver (receiver 2) as side information. In this group, the side information of receivers 1 and 2 about each other's requested messages has this property. Condition \eqref{outer23} is due to the point-to-point channel capacity between the transmitter and receiver 3.
\end{IEEEproof}
\vspace{0pt}
\subsection{Evaluation of the Proposed Inner and Outer Bounds} 
\begin{figure}[t]
\hspace{-18pt}
\includegraphics[width=0.55\textwidth]{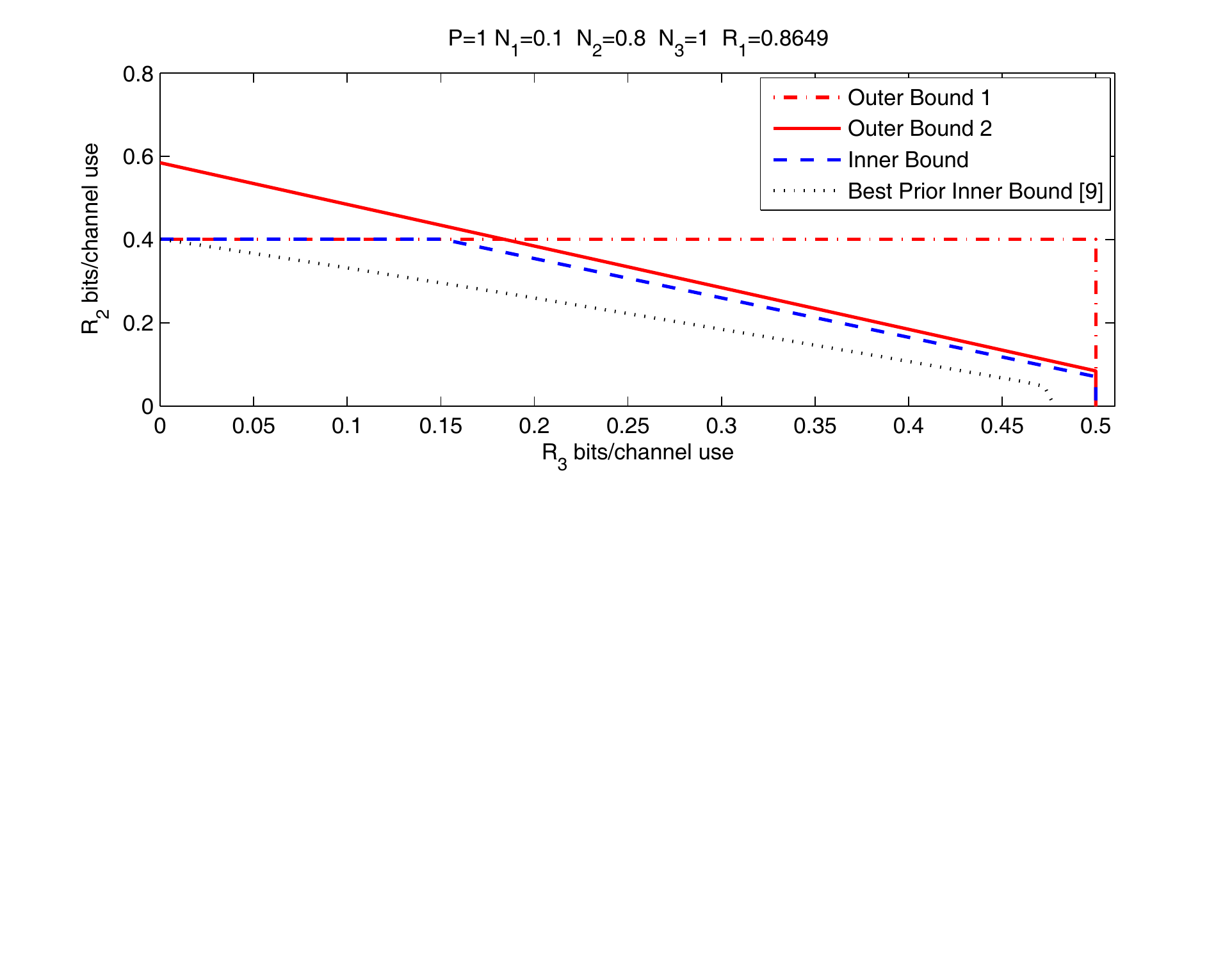}
\vspace{-135pt}
\caption{Inner and outer bounds for $\mathcal{G}_{17}\cup\mathcal{G}_{24}$} 
\vspace{-13pt}
\label{Group7comparison}
\end{figure}

In this subsection, we show that the proposed inner and outer bounds for group 7 coincide for four members and reduce the gap between the best known inner and outer bounds for the remaining four members.

For $\mathcal{G}_{17}\cup\mathcal{G}_{22}$ and $\mathcal{G}_{17}\cup\mathcal{G}_{25}$, the proposed outer bound, $\mathscr{R}'_{\text{out}_1}\cap\mathscr{R}'_{\text{out}_2}$, coincides with $\mathscr{R}'_{\text{in}}$ which consequently establishes the capacity. This is while $\mathscr{R}'_{\text{out}_2}$ (the best existing outer bound) alone is not tight for these members. 

For $\mathcal{G}_{17}\cup\mathcal{G}_{27}$ and $\mathcal{G}_{17}\cup\mathcal{G}_{28}$, $\mathscr{R}'_{\text{out}_1}$, given in \eqref{outer21}--\eqref{outer23}, coincides with $\mathscr{R}'_{\text{in}}$. This establishes the capacity for these members and shows that $\mathscr{R}'_{\text{out}_1}$ is strictly tighter than $\mathscr{R}'_{\text{out}_2}$ for these members (this is because $\mathscr{R}'_{\text{out}_1}$ has some curved surfaces while $\mathscr{R}'_{\text{out}_2}$ is a polyhedron).

For the remaining four members, we show that the inner and outer bounds are both tighter than the best existing ones.

The best existing inner bound \cite{Capacity3UsersPrivateMessage} for the remaining four members with unknown capacity is the set of all rate triples $(R_1,R_2,R_3)$, each satisfying
\begin{align}
R_2\hspace{-2pt}+\hspace{-15pt}\sum_{i\in\{1,3\}\setminus\mathcal{O}_1}{\hspace{-15pt}R_i}&<B_1+B_2+B_3,\label{bestpriorinner1}\\
R_2+R_3&<B_2+B_3,\label{bestpriorinner2}\\
R_3&<\min\{C\left(\frac{\alpha_3P}{N_3}\right),B_3\},\label{bestpriorinner3}
\end{align}
where $B_1\hspace{-4pt}=\hspace{-4pt}C\left(\alpha_1P/{N_1}\right)$, $B_2=C\left({\alpha_2P}/{\left(\alpha_1 P+N_2\right)}\right)$, and $B_3=C\left({\alpha_3P}/{((\alpha_1+\alpha_2)P+N_2)}\right)$ for some $\alpha_k\geq0\;k=1,2,3$ such that $\sum_{k=1}^{3}{\alpha_k}=1$.
This inner bound for $\mathcal{G}_{17}\cup\mathcal{G}_{21}$ and $\mathcal{G}_{17}\cup\mathcal{G}_{23}$ is achieved by using the following scheme
\begin{equation*}
x_1^{(n)}(m_{10})+x_2^{(n)}([m_{11}, m_{20}])+x_3^{(n)}([m_{12},m_{21},m_3]),
\end{equation*}
and for $\mathcal{G}_{17}\hspace{-2pt}\cup\hspace{-2pt}\mathcal{G}_{24}$ and $\mathcal{G}_{17}\hspace{-2pt}\cup\hspace{-2pt}\mathcal{G}_{26}$ by using the following scheme
\begin{equation*}
x_1^{(n)}(m_{10})+x_2^{(n)}([m_{11}, m_{20}])+x_3^{(n)}([m_{12}\hspace{-2pt}\oplus\hspace{-2pt}m_3,m_{21}]),
\end{equation*}
where the three subcodebooks are constructed independently using i.i.d. codewords generated according to $X_k\hspace{-2pt}\sim\hspace{-2pt}\mathcal{N}(0,\alpha_kP)\;k=1,2,3$ for some $\alpha_k\geq0$ such that $\sum_{k=1}^{3}\alpha_k\hspace{-2pt}=\hspace{-2pt}1$. Also, the receivers employ a joint decoding approach \cite{Capacity3UsersPrivateMessage} which utilizes side information during successive decoding. 

For these members, we now show that for any chosen set of $\{\alpha_k\}_{k=1}^3$, the region in \eqref{bestpriorinner1}--\eqref{bestpriorinner3} is smaller than $\mathscr{R}'_{\text{in}}$ for $\alpha=\alpha_1$. Noting that $B_2+B_3=C\left(\frac{(1-\alpha_1)P}{\alpha_1 P+N_2}\right)$, then condition $(a)$ is the same as \eqref{bestpriorinner1}, conditions $(b)$ and $(c)$ are more relaxed than \eqref{bestpriorinner2}, and condition $(d)$ is more relaxed than \eqref{bestpriorinner3}. This proves that the proposed inner bound is larger than best existing inner bound for these members.

Concerning the outer bound, since the proposed outer bound is the intersection of the best existing outer bound and a new outer bound, $\mathscr{R}'_{\text{out}_1}$, the proposed outer bound is tighter than the best existing outer bound. As an example, for $\mathcal{G}_{17}\cup\mathcal{G}_{24}$, Fig. \ref{Group7comparison} depicts that the proposed inner bound is strictly larger than the best existing one, and the proposed outer bound is strictly tighter than the best existing one. 
\vspace{-0pt}
\section{Conclusion}
We considered the problem of private-message broadcasting over the three-receiver AWGN BC with receiver message side information. Following the recent classification of all possible side information configurations into eight groups and the establishment of the capacity for six groups, we investigated the capacity of the channel for the two remaining groups with unknown capacity, groups 4 and 7. We proposed inner and outer bounds for these two groups. For group 4, the proposed inner and outer bounds coincide at certain regions. For group 7, the proposed inner and outer bounds coincide for four members, and for the remaining four members, the proposed inner and outer bounds are both tighter than the best existing ones.
\vspace{-0pt}
\bibliographystyle{IEEEtran}
%

\end{document}